\documentclass[twocolumn,english,pra,showpacs]{revtex4-1}
\usepackage[T1]{fontenc}
\usepackage[latin9]{inputenc}
\usepackage{amsmath}
\usepackage{graphicx}
\usepackage{amssymb}
\usepackage{esint}

\makeatletter
\@ifundefined{textcolor}{}
{%
 \definecolor{BLACK}{gray}{0}
 \definecolor{WHITE}{gray}{1}
 \definecolor{RED}{rgb}{1,0,0}
 \definecolor{GREEN}{rgb}{0,1,0}
 \definecolor{BLUE}{rgb}{0,0,1}
 \definecolor{CYAN}{cmyk}{1,0,0,0}
 \definecolor{MAGENTA}{cmyk}{0,1,0,0}
 \definecolor{YELLOW}{cmyk}{0,0,1,0}
 }

\makeatletter
\usepackage{dcolumn}
\usepackage{bm}
\usepackage{amsthm}\@ifundefined{definecolor}
 {\@ifundefined{definecolor}
 {\usepackage{color}}{}
}{}

\makeatother

\usepackage{babel}

\makeatother

\usepackage{babel}

\begin{document}
\newtheorem{conjecture}{Conjecture}\newtheorem{corollary}{Corollary}\newtheorem{theorem}{Theorem}
\newtheorem{lemma}{Lemma}
\newtheorem{observation}{Observation}
\newtheorem{definition}{Definition}\newtheorem{remark}{Remark}\global\long\global\long\def\ket#1{|#1 \rangle}
\global\long\global\long\def\bra#1{\langle#1|}
\global\long\global\long\def\proj#1{\ket{#1}\bra{#1}}

\title{Optimal work of quantum Szilard engine}

\author{Hee Joon Jeon}
\affiliation{Department of Physics, Pusan National University, Busan 609-735 Korea}

\author{Sang Wook Kim}
\affiliation{Department of Physics Education, Pusan National University, Busan 609-735 Korea }

\date{\today}


\begin{abstract}
We have found the optimal condition of the work performed by the quantum Szilard engine (SZE) containing multi-particles. Usually the optimal work of a cyclic engine is achieved when the whole thermodynamic process is reversible. Although the quantum SZE inherently contains an irreversible process, we can still define effectively reversible protocol based upon the force of time-forward and time-backward processes.
\end{abstract}

\maketitle

Maxwell was the first man that asked a deep physical question on the nature of information using his famous demon \cite{Leff03}. Szilard then recognized the connection between information and entropy, which is demonstrated by his thought experiment called as the Szilard engine (SZE) \cite{Szilard29}. It tells us that one can extract work from a cyclic engine with a single heat bath by exploiting information. It seems to violate Carnot's version of the second law of thermodynamics. However, it is now widely accepted that the SZE does not violate the second law, but the Carnot's principle is necessary to modify \cite{Maruyama09}. The SZE has been realized in various experiments \cite{Serreli07,Price08,Thorn08,Toyabe10}.

\begin{figure}[]
\includegraphics[width=0.5\textwidth]{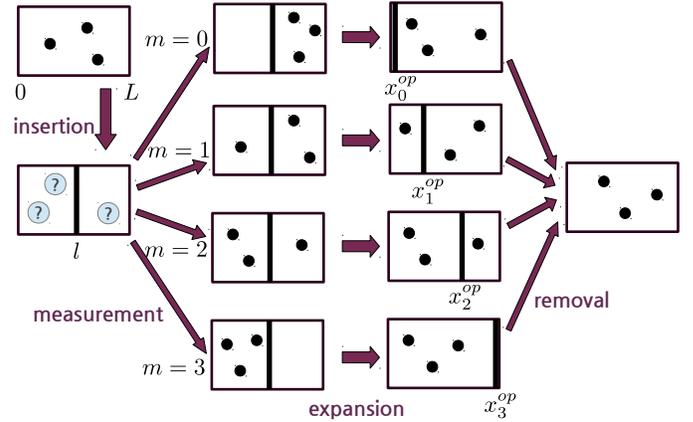}
\caption{Schematic diagram of the thermodynamic processes of the quantum SZE. (See the text for the details.)}
\label{fig1}
\end{figure}

The thermodynamic cycle of the SZE containing $N$ particles in the one-dimensional box of the length $L$ consists of four stages as shown in Fig.~\ref{fig1}; (i) to insert a wall at $x=l$ $(0 \leq l \leq L)$, (ii) to measure how many particles are in the left side, whose outcome is expressed as $m$, (iii) to perform isothermal expansion, where the wall stops at $x=x_m^0$ differing in $m$, and (iv) to remove the wall to complete the cycle. It has been found that the quantum mechanical work done by the SZE is given by \cite{SW_Kim11}
\begin{equation}
\label{eq_work_tot}
W = -k_BT \sum_{m=0}^N f_m(l) \ln \left[ {f_m(l) \over f_m^\star(x_m) }  \right]
\end{equation}
with $f_m(l) = Z_m(l) / Z(l) $, $f_m^\star(x_m) = Z_m(x_m) / Z(x_m)$ and $Z(y) = \sum_m Z_m(y)$. Here $Z_m(y)$ denotes the partition function of the case that $m$ particles are in the left side of the wall, i.e. $N-m$ particles in the right, when the wall is located at $y$. Note that $l$ is freely chosen when the wall is inserted, while $x_m^0$ is determined from the force balance, $F^{\rm L}+F^{\rm R}=0$. The generalized force of the left (right) side of the wall is defined as
\begin{equation}
F^{\rm L(R)} = \sum_n P_n^{\rm L(R)} \left[\frac{\partial E_n^{\rm L(R)}}{\partial x}\right],
\label{eq_gen_force}
\end{equation}
where $P_n^{\rm L(R)}$ and $E_n^{\rm L(R)}$ represent the occupation probability and the eigenenergy of the $n$th energy level of the left (right) side, respectively.

The work of the {\em classical} SZE is also expressed as Eq.~(\ref{eq_work_tot}), which is derived by using the dissipative work theorem of {\em classical} non-equilibrium thermodynamics \cite{KH_Kim11}. The reason why the non-equilibrium thermodynamics is taken into account is that it is an irreversible process that the wall separating particles into two parts is removed. This is nothing but free expansion. Even though both quantum and classical SZE have the equivalent mathematical form of the work, Eq.~(\ref{eq_work_tot}), the partition functions differ in classical and quantum mechanics so that the amount of work of the quantum SZE is distinct from that of the classical one \cite{KH_Kim11}. Note that information itself is inherently classical both in the quantum \cite{SW_Kim11} and in the classical SZE \cite{KH_Kim11}. Recently the heat engine exploiting purely quantum information is proposed \cite{Park13}.

\begin{figure}[]
\includegraphics[width=0.5\textwidth]{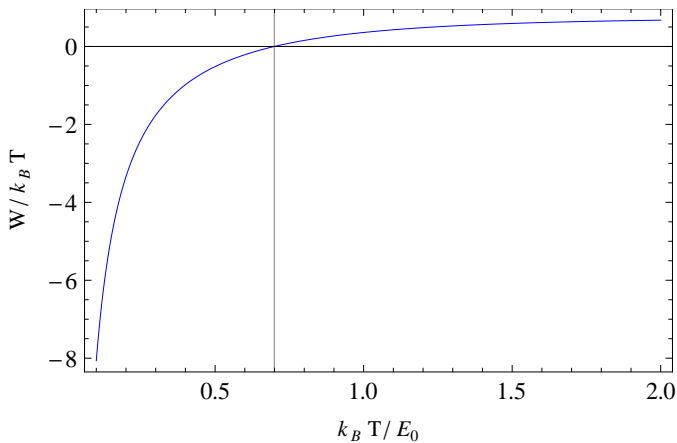}
\caption{The work of the quantum SZE containing three bosons with $l=L/2$ as a function of temperature when the wall stops at $x_m^0$ (the force balance point). For low temperature the work becomes negative.}
\label{fig2}
\end{figure}

Strangely enough in the quantum SZE, however, it has been found that the work of Eq.~(\ref{eq_work_tot}) in the low-temperature limit can be negative as shown in Fig.~\ref{fig2}, implying one should do work on the engine instead of extracting work from it \cite{Plesch13}. Later it is found that it is not optimal to stop the wall at the force balance originally proposed in Ref.~\cite{SW_Kim13}. Instead one can numerically find the optimal condition. In a usual heat engine, the optimal condition of work gain is achieved when the whole thermodynamic process is reversible. As far as the quantum SZE is concerned, however, it is unlikely to make it fully reversible since the stage of removing the wall is inherently irreversible as mentioned above. It raises a question on what the physical meaning of the optimal condition of the quantum SZE is.

In this paper we show the optimal work gain of the quantum SZE is achieved if we stop and remove the wall at the location where the force on the wall of the time-forward process is equivalent to the {\em average} force right after inserting the wall for the {\em time-backward} process. Note that the time-backward process of removing the wall is simply inserting the wall. For the time-backward process after inserting the wall one cannot automatically return to the original state of the time-forward process, i.e. $m$ particles in the left, since it is an irreversible process to remove the wall in the time-forward process. Thus it is a reasonable way to define the force of the time-backward as the average over all possible cases with different $m$. Briefly speaking, the optimal condition is then achieved when the time-forward force is equivalent to the time-backward force. In some sense this effectively recovers the reversibility of the inherently irreversible heat engine not because the process itself is reversible but because the forces are reversible. It gives us a new insight on the optimal condition of {\em irreversible} heat engines.

The work of Eq.~(\ref{eq_work_tot}) is a function of $l$ and $\{x_m\}$ with $m=0,1, \cdots, N$, namely $W(l,\{x_m\})$. The optimal (maximum) condition is thus obtained from $\partial W / \partial l = 0$ and $\partial W / \partial x_m = 0$ \cite{comment}. From Eq.~(\ref{eq_work_tot}) the latter is shown to be ${\partial \ln f_m^\star(x_m)} / {\partial x_m}=0$, which is also rewritten as
\begin{equation}
\label{eq_extreme_condition}
{ 1 \over Z_m }{{\partial Z_m} \over {\partial x_m}} = { 1 \over Z } {{\partial Z} \over {\partial x_m}}
\end{equation}
with $Z(y) = \sum_m Z_m(y)$. The partition function $Z_m$ is given as
\begin{equation}
\label{eq_partition_func}
Z_m(y) = \sum_{\sigma} e^{-\beta \epsilon^\sigma_m(y)}
\end{equation}
with $\epsilon^\sigma_m = \sum_n (E^L_n m^\sigma_n + E^R_n q^\sigma_n )$, where $m^\sigma_n$ and $q^\sigma_n$ represent the number of particles occupying the $n$th energy level of the left and the right side, respectively, satisfying $m = \sum_n m^\sigma_n$ and $N-m = \sum_n q^\sigma_n$. Here $\sigma$ denotes the configuration of particles over all the energy levels for a given $m$.

The left-hand side of Eq.~(\ref{eq_extreme_condition}) represents the net force  on the wall multiplied by the inverse temperature when $m$ particles are in the left side due to
\begin{equation}
\label{eq_F_m}
{ 1 \over Z_m }{{\partial Z_m} \over {\partial x_m}} = - \beta \left\langle  {{\partial \epsilon^\sigma_m} \over {\partial x_m}} \right\rangle_\sigma \equiv \beta F_m
\end{equation}
with $\beta = 1/k_B T$. Note that $F_m(x_m^0)=0$ gives rise to the force balance condition. On the other hand, the right-hand side of Eq.~(\ref{eq_extreme_condition}) is expressed as
\begin{equation}
\label{eq_F_avg}
{ 1 \over Z } {{\partial Z} \over {\partial x_m}} = { 1 \over Z } \sum_p {{\partial Z_p} \over {\partial x_m}} = -\beta \sum_p { Z_p \over Z } \left\langle  {{\partial \epsilon^\sigma_p} \over {\partial x_m}} \right\rangle_\sigma \equiv \beta \langle F_p \rangle_p.
\end{equation}
The force $\langle F_p \rangle_p$ of Eq.(\ref{eq_F_avg}) is obtained from averaging $F_p$. It is emphasized that every $F_p$ is calculated with the equivalent $x_m$ at which $F_m$ of Eq.~(\ref{eq_F_m}) is obtained. To understand the physical meaning of this force let us consider the time-forward and back-ward process of removing the wall in detail. Note that the time-backward process of removing the wall is inserting it. For the time-forward the number of particles on the left, $m$, is determined once the measurement is performed, and then is kept constant before we stop the wall at $x_m$ and remove it. For the time-backward, however, one finds $m$ to vary from $0$ to $N$ when the wall is inserted at $x_m$. $\langle F_p \rangle_p$ is regarded as the legitimate average force at $x_m$ for the time-backward since we inevitably confront such a situation when we perform the reverse thermodynamic process of the quantum SZE.

From Eqs.~(\ref{eq_F_m}) and (\ref{eq_F_avg}), the optimal condition Eq.~(\ref{eq_extreme_condition}) reads
\begin{equation}
\label{eq_main}
F_m(x_m^{\rm op})  = \langle F_p(x_m^{\rm op}) \rangle_p.
\end{equation}
The optimal $x_m^{\rm op}$ is obtained from the condition that the force of the time-forward is equivalent to that of the time-backward. It gives an insight on the optimal condition of the heat engine containing an irreversible process. Usually the optimal work of a heat engine is attained if the whole thermodynamic process is performed in a reversible way. Equation (\ref{eq_main}) tells us that the {\em thermodynamically irreversible} quantum SZE becomes optimal if the process is {\em effectively} reversible, i.e. the forces are reversible. The force is important since the work, which we intend to make optimal, is given by integrating force over coordinate.

\begin{figure}[]
\includegraphics[width=0.5\textwidth]{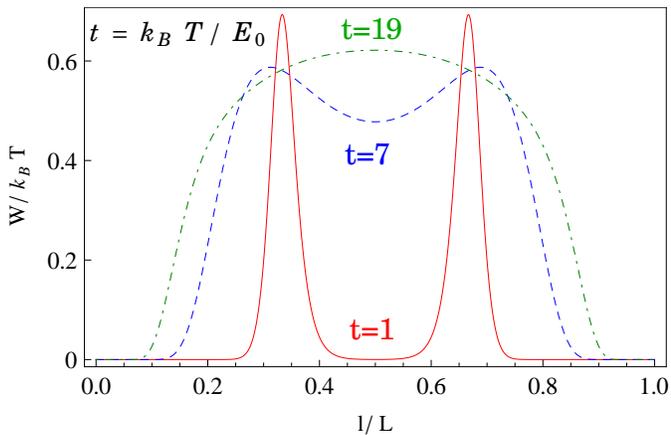}
\caption{The work $W$ of the SZE containing two spinless fermions as a function of $l$ with various $T$.}
\label{fig3}
\end{figure}

The work at the optimal condition is positive. It is recalled that the observation that the work obtained from the force balance, $F_m(x_m^0)=0$, can become negative in the low temperature initiates this research. Mathematically $f_m$ is equivalent to $f_m^\star$ in their form, namely $f_m(y)=f_m^\star(y)$. Since the optimal work is obtained from maximizing $f_m^\star$ [See Eq.~(\ref{eq_work_tot})], i.e. ${\partial \ln f_m^\star(x_m)} / {\partial x_m}=0$, we find  $f_m(l) = f_m^\star(l) \le f_m^\star(x_m^{\rm op})$ irrespective of $l$. It immediately implies that the work of Eq.~(\ref{eq_work_tot}) at maximum is positive.

In fact, it is the quantum effect that $x_m^0$ of the force balance differs from $x_m^{\rm op}$ of the optimal condition. If we consider classical particles satisfying ideal gas law implying pressure is proportional to density, one finds
\begin{equation}
\label{eq_F_avg_cl}
\langle F_p(y) \rangle_p = \sum_{p=0}^N P(p) \left( \frac{p}{y} - \frac{N-p}{L-y} \right) = 0
\end{equation}
for any $y \in (0,L)$ with
\begin{equation}
P(p)=\left( \frac{y}{L} \right)^p \left( \frac{L-y}{L} \right)^{N-p} \begin{pmatrix} N  \\ p \end{pmatrix}.
\end{equation}
The optimal condition is thus achieved simply by the force balance since the right-hand side of Eq.~(\ref{eq_main}) always vanishes. Note that $x_m^0$ of the classical particles are different from that of the quantum ones; They coincide only in the high temperature limit. This will be discussed below in detail.

\begin{figure}[]
\begin{center}$
\begin{array}{cc}
\includegraphics[width=0.5\textwidth]{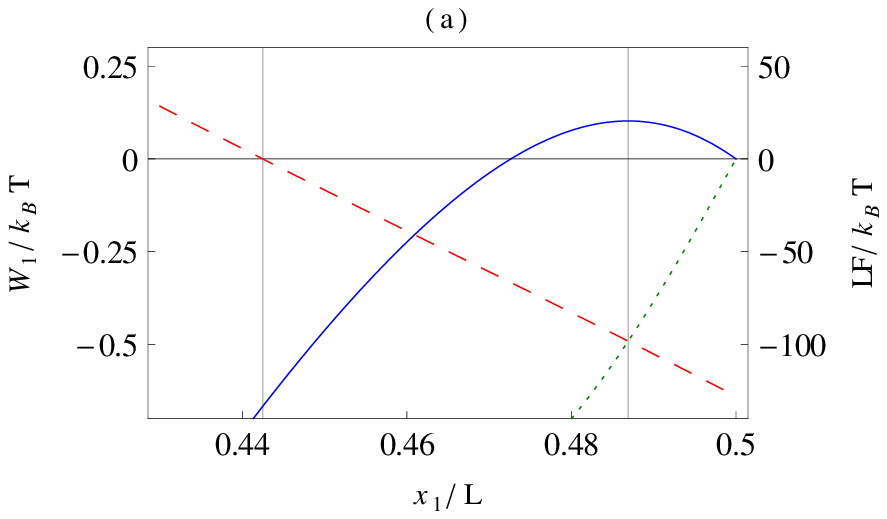} \\
\includegraphics[width=0.5\textwidth]{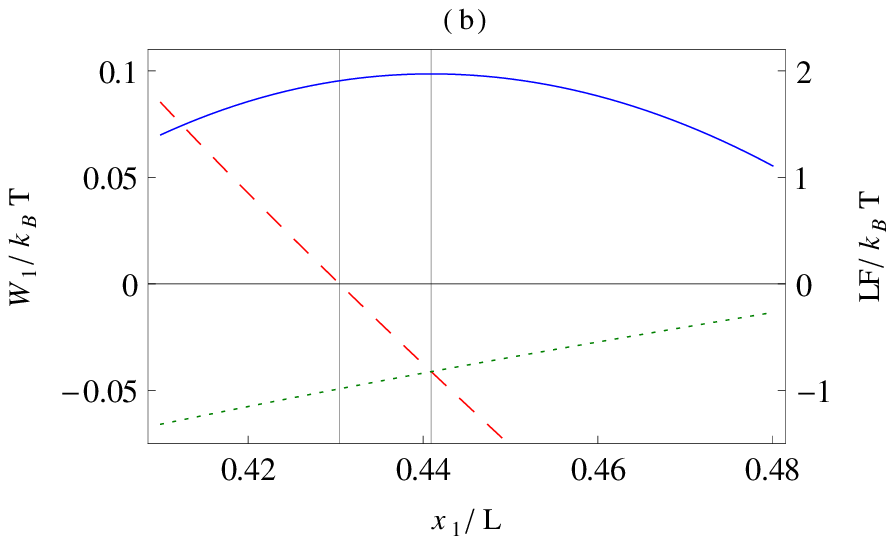}
\end{array}$
\end{center}
\caption{The partial work $W_1(L/2,x_1)$ (the solid curve; see the left hand side of $y$ axis) of the SZE containing three bosons as a function of $x_1$ for (a) $k_BT/E_0=1$ and (b) $k_BT/E_0=5$. The dashed and the dotted curves represent $F_1(x_1)$ and $\langle F_p(x_1) \rangle_p$ (see the right hand side of $y$ axis), respectively. The former crosses $x$ axis at $x_1^0$ denoted by the left vertical line, and intersects the latter at $x_1^{\rm op}$ denoted by the right vertical line.}
\label{fig4}
\end{figure}

In some sense it looks weird to have $\langle F_p(x_m^{\rm op}) \rangle_p \neq 0$ in quantum mechanics. This is simply the net force exerted on a wall when we insert the wall into a box containing particles kept in equilibrium during the process in time-backward process. The non-zero net force cannot be spontaneously developed as far as only quasi-static process of inserting the wall is concerned in classical thermodynamics. We emphasize that it is the quantum effect. Interestingly this seems to be similar to the Casimir effect \cite{Hertzberg05}. If the wall is inserted at $x \neq L/2$, the structures of energy levels of the left and the right sides become different so do the corresponding generalized forces of Eq.~(\ref{eq_gen_force}), namely $\sum_n P_n^{\rm L} ({\partial E_n^{\rm L}}/{\partial x})$ and $\sum_n P_n^{\rm R} ({\partial E_n^{\rm R}}/{\partial x})$.

So far we have discussed the optimal condition of $\{x_m\}$. The optimal condition of $l$ is rather subtle. Intuitively one expects the symmetric point $l=L/2$ satisfies the optimal condition. At least we can show that the work (\ref{eq_work_tot}) has an extremum at $l=L/2$. In general, however, the work does not always exhibit a global maximum at $l=L/2$. We set $x_m=x_m^{\rm op}$ since $x_m^{\rm op}$ has nothing to do with $l$. Then we obtain from $\partial W / \partial l = 0$
\begin{equation}
\label{eq_opt_l}
\langle W_m(l,x_m^{\rm op}) F_m(l) \rangle_m = \langle W_p(l,x_p^{\rm op}) \rangle_p \langle F_q(l) \rangle_q
\end{equation}
with $W_m(l,y)= - \ln \left[ f_m(l) / f^\star_m(y) \right]$. Both the left and the right hand sides of Eq.~(\ref{eq_opt_l}) vanishes at $l=L/2$ according to $f_{N-m}(L/2)=f_m(L/2)$, $W_{N-m}(L/2,x_{N-m}^{\rm op})=W_m(L/2,x_m^{\rm op})$ and  $F_{N-m}(L/2)=-F_m(L/2)$, so that the work exhibits an extremum at $l=L/2$. However, this is not a unique solution. For example, Figure~\ref{fig3} shows that a single maximum of the work of the SZE containing two spinless fermions splits into two peaks as temperature decreases although the work remains extremum at $l=L/2$. This splitting is associated with the accidental degeneracies of the problem and the exclusion principle of fermions \cite{KH_Kim12}. The optimal condition of the work of $l$ thus depends on temperature and the number of particles in fermions, so that it is not easy to find its simple mathematical expression. However, we set $l=L/2$ below since it is the optimal condition for fermions except extremely low temperature and bosons. Moreover, the inherent irreversibility of the SZE is associated with $x_m$, at which the wall is removed, rather than $l$.

\begin{figure}[]
\includegraphics[width=0.5\textwidth]{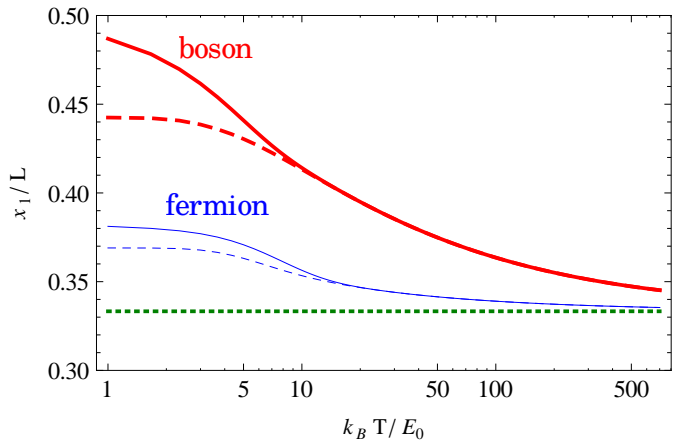}
\caption{The optimal $x_1^{\rm op}$ of the SZE of three bosons (the thick solid curve) and three fermions (the thin solid curve), and the force balance point $x_1^0$ of three bosons (the thick dashed curve) and three fermions (the thin dashed curve) as a function of temperature. The horizontal dotted line represents the optimal condition of three classical particles, $L/3$ (temperature-independent).}
\label{fig5}
\end{figure}

Now let us consider the quantum SZE containing three particles as an example. The wall is inserted at $L/2$. The box is described by the infinite potential well. Recall that the optimal work is obtained when each $f_m^\star(x_m)$ is maximized. For $m=0$ and $m=N$ the optimal condition is trivially achieved when $f_0^\star=f_N^\star=1$ with $x_0=0$ and $x_N=L$, respectively. Moreover, due to the symmetry the situation with $m=1$ is equivalent to that with $m=2$, so that it is enough to consider the work of only $m=1$ of Eq.~(\ref{eq_work_tot}), namely $W_1(L/2,x_1)$. Figure \ref{fig4}(a) shows that in the low temperature $W_1$ of the quantum SZE containing three bosons is negative at the force balance, $x_1 =x_1^0 \approx 0.443$. However, the {\em positive} optimal work is obtained at $x_1 = x_1^{\rm op} \approx 0.490$ where two curves $F_1(x_1)$ and $\langle F_p(x_1) \rangle_p$ are crossed. It is shown in Fig.~\ref{fig4}(b) that for rather higher temperature $W_1$ is not only positive but $x_1^{\rm op}$ also becomes closer to $x_1^0$. The reason is that quantum partition functions become equivalent to the classical ones, so do $f_m^\star$'s, which implies $\langle F_p(y) \rangle_p \rightarrow 0$ as $T \rightarrow \infty$ according to Eq.~(\ref{eq_F_avg_cl}). Figure \ref{fig5} shows for both three bosons and three fermions $x_1^{\rm op}$ approaches $x_1^0$ and finally converges to the classical result, $L/3$, as temperature increases.

In summary, we have found the physical meaning of the optimal condition of the work done by quantum SZE containing multi-particles and shown that the optimal work is positive. The optimal condition is satisfied if the wall is stopped and removed when the force on the wall of time-forward process is equivalent to that of time-backward. It sheds some light on the optimal condition of the irreversible information heat engine.

This research was supported by Basic Science Research Program through the National Research Foundation of Korea(NRF) funded by the Ministry of Science, ICT and future Planning (2009-0087261 and 2013R1A1A2011438).


\begin{thebibliography}{00}
\bibitem{comment}
In fact, these guarantee that they are only extrema. Instead of finding their second derivatives, we have confirmed they are maxima by directly plotting them like Fig.~\ref{fig3}.
\end{thebibliography}
\end{document}